\documentclass[12pt]{article}
\usepackage{amssymb,amsmath}
\usepackage{epsfig}

\newcommand{\sect}[1]{\setcounter{equation}{0}\section{#1}}

\def\N{{\mathcal N}}

\def\O{{\mathcal O}}
\def\ds{\displaystyle}

\def\a{\alpha}
\def\b{\beta}

\def\g{\gamma}

\def\s{\sigma}
\def\t{\tilde}

\def\m{\mu}
\def\n{\nu}

\def\vf{\varphi}

\def\l{\lambda}
\def\th{\theta}

\def\o{\omega}
\def\p{\partial}

\def\axs{AdS_5\times S^5}
\newcommand{\eq}[1]{\begin{equation} #1 \end{equation}}

\begin{document}

\begin{center}

\baselineskip=18pt

\vspace*{2cm}

{\bf{\Large Multispin Giant Magnons}  \\
\vspace*{.35cm} }

\vspace*{1cm} N.P. Bobev${}^{\star}$ and R.C.
Rashkov${}^{\dagger}$\footnote{e-mail: rash@phys.uni-sofia.bg;
bobev@usc.edu}

\ \\
${}^{\star}$ \textit{Department of Physics and Astronomy ,
University of Southern California, Los Angeles, CA 90089-0484,
USA}

\ \\

${}^{\dagger}$ \textit{Department of Physics, Sofia University,
1164 Sofia, Bulgaria}

\end{center}

\vspace*{.8cm}

\baselineskip=18pt

\begin{abstract}
We investigate giant magnons from classical rotating strings in
two different backgrounds. First we generalize the solution of
Hofman and Maldacena and investigate new magnon excitations of a
spin chain which are dual to a string on $R\times S^5$ with two
non-vanishing angular momenta. Allowing string dynamics along the
third angle in the five sphere, we find a dispersion relation that
reproduces the Hofman and Maldacena one and the one found by Dorey
for the two spin case. In the second part of the paper we
generalize the two "spin" giant magnon to the case of
$\b$-deformed $\axs$ background. We find agreement between the
dispersion relation of the rotating string and the proposed
dispersion relation of the magnon bound state on the spin chain.

\end{abstract}

\vspace*{.8cm}

\newpage \baselineskip=18pt

\sect{Introduction}

The AdS/CFT correspondence \cite{ads/cft} is a powerful tool for
studies and future advances in the understanding of gauge theories
at strong coupling. One of the predictions of the correspondence
is the equivalence between the spectrum of free string theory on
$\axs$ and the spectrum of anomalous dimensions of gauge invariant
operators in the planar $\N=4$ Supersymmetric Yang-Mills (SYM)
theory. In order to check this conjecture we need the full
spectrum of free string theory on a curved space, such as $\axs$,
which is still an unsolved problem. However in certain limits the
problem becomes tractable and some comparisons on both side of the
correspondence can be made. It turns out that the problem
simplifies in the limit of string states with large quantum
numbers \cite{bmn},\cite{GKP} and one can use a semiclassical
approximation on the string side to find the energy spectrum. In
order to test the predictions of the correspondence a way to
compute the anomalous dimensions of long single trace operators
(dual to the strings with large quantum numbers) on the gauge
theory side was needed. Minahan and Zarembo proposed a remarkable
solution to this problem \cite{mz} by relating the Hamiltonian of
the Heisenberg spin chain with the dilatation operator of $\N=4$
SYM. On other hand, in several papers the relation between strings
and spin chains was established, see for instance
\cite{kruczenski},\cite{dim-rash},\cite{ lopez},\cite{tseytlin1}
and references therein. This idea opened the way for a remarkable
interplay between spin chains, gauge theories, string theory
\footnote{For very nice reviews on the subject with a complete
list of references see
\cite{beisphd},\cite{tseytrev},\cite{Plefka:2005bk}} and
integrability (the integrability of classical strings on $\axs$
was proven in \cite{bpr}).

Having the above relations, it was natural to look for string
solutions governing various corners of the spin chain spectrum.
The most studied cases were spin waves in long-wave approximation
corresponding to rotating and pulsating strings in certain limits,
see for instance reviews
\cite{beisphd},\cite{tseytrev},\cite{Plefka:2005bk} and references
therein. Another interesting case are the low lying spin chain
states corresponding to the magnon excitations. One class of
string solutions already presented in a number of papers is the
string theory on pp-wave backgrounds. The later, although
interesting and important, describe point-like strings which are
only part of the whole picture. The question of more general
string solutions corresponding to this part of the spectrum was
still unsolved.

Recently Maldacena and Hofman\cite{HM} were able to map spin chain
"magnon" states to specific rotating semiclassical string states
on $R\times S^2$. This result was soon generalized to magnon bound
states (\cite{Dorey1},\cite{Dorey2},\cite{AFZ},\cite{MTT}), dual
to strings on $R\times S^3$ with two non-vanishing angular
momenta. Moreover in \cite{MTT} a different giant magnon state
with two spins was found, it is dual to string moving on
$AdS_3\times S^1$ i.e. it has spin in both the AdS and the
spherical part of the background. It is natural to generalize this
construction and look for giant magnons with multiple spins, these
should be dual to rotating strings on $R\times S^5$ with 3 spins.

The relation between energy and angular momentum for the one spin
giant magnon found in \cite{HM} is:

\eq{ E-J=\ds\frac{\sqrt{\l}}{\pi}|\sin\ds\frac{p}{2}| }

where $p$ is the magnon momentum which on the string side is
interpreted as a difference in the angle $\phi$ (see \cite{HM} for
details). In the two spin case the $E-J$ relation was found both
on the string \cite{Dorey2},\cite{AFZ},\cite{MTT} and spin chain
\cite{Dorey1} sides and looks like:

\eq{ E-J=\ds\sqrt{J_2^2+\ds\frac{\l}{\pi^2}\sin^2\frac{p}{2}} }

where $J_2$ is the second spin of the string. Allowing analogous
string dynamics on the third angle of the five sphere one can ask
what is the dispersion relation and to which magnon excitations it
should correspond. For for the particular string configuration we
consider in the first part of the paper we find agreement with two
spin magnon excitations on the spin chain side\footnote{We thank
A.A. Tseytlin for comments on this point}.

In the second part of the paper we will investigate multi spin
giant magnons in the $\b$-deformed $\axs$ background found by
Lunin and Maldacena \cite{LM} \footnote{For related work on
string/gauge/spin chain duality in $\b$-deformed backgrounds see
\cite{Frolov}-\cite{Gursoy:2006gm}}. We will generalize the
results in \cite{khoze} by considering a more general rotating
string anzatz. The same anzatz was considered in
\cite{AFZ},\cite{MTT} for the pure $\axs$ case. It was argued in
\cite{Dorey1} that this simple rotating string with two angular
momenta corresponds to a bound state of magnons propagating on the
spin chain. The second angular momentum $J_2$ was interpreted as
the number of magnons in this bound state (this number can be
finite). As expected in the limit of vanishing deformation -
$\b\rightarrow 0$ we reproduce the dispersion relation from the
pure $\axs$ case.

The paper is organized as follows: In section 2 we briefly
describe the Hofman-Maldacena giant magnon solution from strings
on $R \times S^2$ and its recently found two spin generalization.
We then proceed with applying the same idea to rotating strings on
$R\times S^5$ and find the energy-momentum relation for these
strings. In section three we turn our attention to the so called
$\b$-deformed $\axs$ background and investigate two spin giant
magnons in it. This is a generalization of the two spin giant
magnon presented in section 2 and is different from the solution
found in \cite{khoze}. In the last section we present our
conclusions and a few possible directions for further studies.


\section{Giant magnons from strings on $\axs$}

Here we will review the idea of Hofman and Maldacena \cite{HM} for
the stringy description of spin chain magnon excitations.  Let us
look of a special class of gauge invariant operators which are
characterized by an infinite value of one of the $SO(6)$ charges
$J$ and have finite value of $E-J$. When $E-J$ vanishes the
operator is composed of a chain of $J$ operators, namely:

\eq{\O\sim Tr(ZZ..Z...Z)}

On this "chain" of we can consider a finite number of fields $Y$
which can be interpreted as excitations propagating on the chain.
The corresponding operator takes the form:

\eq{\O\sim \ds\sum_{k}e^{ikp}(Z..ZZYZZ...Z)}

Where the summation is over all possible positions on the chain
where $Y$ can be inserted. This picture on the gauge side has a
description in terms of spin chains \cite{mz}. The operators $Z$
represent the ground state of the chain and the $Y$s correspond to
magnon excitations.

Using supersymmetry arguments Beisert \cite{beisert} was able to
find the following dispersion relation of the magnon excitations:

\eq{ E-J=\sqrt{1+\ds\frac{\l}{\pi^2}\sin^2\ds\frac{p}{2}}
\label{2.1}}

Here $p$ is the magnon momentum and the periodicity in $p$ is due
to the discreteness of the spin chain. The large 't Hooft coupling
limit of the above relation is:

\eq{E-J=\ds\frac{\sqrt{\l}}{\pi}\mid\sin\ds\frac{p}{2}\mid
\label{2.2}}

Hofman and Maldacena were able to reproduce the above relation
from string theory on a $R\times S^2$ subspace of $\axs$. They
considered a simple string with both ends rotating on the equator
of $S^2$ with constant separation between the ends
$\Delta\varphi$. The crucial step was to identify this angle
difference with the momentum of the spin chain magnon
$\Delta\varphi=p$. In order to be able to reproduce exactly the
magnon dispersion relation the following limits are considered:

\eq{J\rightarrow\infty \qquad E-J={\mathrm{fixed}} \qquad
\l={\mathrm{fixed}} \qquad p={\mathrm{fixed}}}

 We note that a closed folded string
configuration with one angular momentum as the one considered in
\cite{GKP} will correspond to a state which is a superposition of
two magnons with opposite momenta. The solution we are interested
in lives on the $R\times S^2$ background with the metric:

\eq{ds^2=-dt^2+d\theta^2+\sin^2\theta d\varphi^2\label{2.3}}

We look for a rotating string solution of the Polyakov action of
this model of the form:

\eq{ t=\tau \qquad \theta=\theta(y) \qquad
\varphi=t+\t{\varphi}(y)\label{2.4}}

Where we have defined the variable $y=cx-dt$. Here $x$ is a world
sheet spatial coordinate with infinite range and $c$ and $d$ are
constants. Using the above anzatz it is not hard to integrate the
equations of motion and get:

\eq{ \cos\theta=\ds\frac{\cos\theta_0}{\cosh y} \qquad
\tan\t{\varphi}=\ds\frac{\tanh y}{\tan\theta_0} \label{2.5}}

The constants $c$ and $d$ are related to the integration constant
$\theta_0$ by:

\eq{\ds\frac{1}{c}=\cos\theta_0 \qquad\qquad
d=\tan\theta_0\label{2.6}}

Now it is easy to see that $c^2-d^2=1$, this provides us with the
necessary condition that the group velocity $v=\frac{d}{c}$ is
less than one. We are interested in the conserved charges for our
system:

\eq{E=\ds\frac{\sqrt{\l}}{2\pi}\ds\int_{-\infty}^{\infty}dx \qquad
J=\ds\frac{\sqrt{\l}}{2\pi}\ds\int_{-\infty}^{\infty}dx\sin^2
\theta \label{2.7}}

Using the above relation we see that

\eq{E-J=\ds\frac{\sqrt{\l}}{\pi}\cos\theta_0 \label{2.8}}

Now the proposal of Hofman and Maldacena \cite{HM} is to make the
identification $\cos\theta_0=\sin(p/2)$. This leads to the
conclusion that the dispersion relations of the magnon excitation
of the infinte spin chain in the large 't Hooft coupling limit and
the rotating string with one large angular momentum are the same.
Thus we have outlined the idea in \cite{HM} for finding the string
dual description of a giant magnon excitation in a certain limit.
This opens the possibility for generalizing this construction and
studying multi magnon bound states, which are argued to be dual
\cite{Dorey1} to rotating strings with two angular momenta. We
proceed with this analysis in the following sections.


\subsection{Two spin giant magnons in $\axs$}

Here we will present the two spin rotating string solution found
in \cite{AFZ} in a slightly different notation which will be more
convenient when we analyze the $\g$-deformed background. This
rotating string is dual to a bound state of $J_2$ magnons with
charge $J_1$. We look for classical string solution on a $R\times
S^3$ subspace of $\axs$:

\eq{ ds^2=-dt^2+d\theta^2+\sin^2\theta d\phi_1^2+\cos^2\theta
d\phi_2^2\label{2.9}}

The Polyakov action for this system is:

\eq{ S_P=\ds\frac{\sqrt{\l}}{4\pi}\int d\tau d\s [-(\p_{\tau}t)^2-(\p_{\tau}\theta)^2+(\p_{\s}\theta)^2+\sin^2\theta((\p_{\s}\phi_1)^2-(\p_{\tau}\phi_1)^2)+\\\\
\cos^2\theta((\p_{\s}\phi_2)^2-(\p_{\tau}\phi_2)^2)] \label{2.10}}

It can be checked that the following anzatz is consistent with the
equations of motion following from the above action and the
Virasoro constraints:

\eq{\theta=\theta(y) \qquad \phi_1=t+g_1(y) \qquad \phi_2=\n
t+g_2(y)\label{2.11}}

Here we have defined a new variable $y=c\s-d\tau$ and we work in
the conformal gauge $t=\tau$. It is easy to derive the following
equations for the unknown functions $g_1(y)$,$g_2(y)$ and
$\theta(y)$:

\eq{\begin{array}{l} \p_yg_1=\ds\frac{d}{\sin^2\theta}-d \qquad
\p_yg_2=-\n d\\\\
\p_y\theta=\cos\theta\ds\sqrt{1-\n^2
c^2-d^2\ds\frac{\cos^2\theta}{\sin^2\theta}}
\end{array}\label{2.12}}

We can integrate the last equation and find $\cos\theta$ to be:

\eq{
\cos\theta=\sqrt{\ds\frac{1-\n^2c^2}{c^2-\n^2c^2}}\ds\frac{1}{\cosh(\sqrt{1-\n^2
c^2}y)} \label{2.13}}

We denote

\eq{ \cos\theta_0=\sqrt{\ds\frac{1-\n^2c^2}{c^2-\n^2c^2}}
\label{2.14}}

and following \cite{HM} we identify $\sin\frac{p}{2}$ with
$\cos\theta_0$, where $p$ is the momentum of the magnon
(interpreted as a geometrical angle in the string picture).
Knowing the solution $\theta(y)$ we can find explicitly the
conserved charges of the rotating string:

\eq{\begin{array}{l}
E=\ds\frac{\sqrt{\l}}{2\pi}\ds\int_{-\infty}^{\infty}d\s\\\\
J_1=\ds\frac{\sqrt{\l}}{2\pi}\ds\int_{-\infty}^{\infty}d\s
\sin^2\theta\left(1-d\p_yg_1\right)\\\\
J_2=\ds\frac{\sqrt{\l}}{2\pi}\ds\int_{-\infty}^{\infty}d\s
\cos^2\theta\left(\n-d\p_yg_2\right)
\end{array}\label{2.15}}

After a little algebra we arrive at:

\eq{
E-J_1=\ds\frac{J_2}{\n}=\ds\frac{\sqrt{\l}}{\pi}\ds\frac{c\cos^2\theta_0}{\sqrt{1-\n^2c^2}}
\label{2.16}}

Finally we extract the dispersion relation for the magnon bound
state found in \cite{Dorey1} and derived from the string sigma
model on $R\times S^3$ in \cite{AFZ},\cite{MTT}:

\eq{E-J_1=\sqrt{J_2^2+\ds\frac{\l}{\pi^2}\sin^2\frac{p}{2}}\label{2.17}}


\subsection{More two spin giant magnons on $R\times S^5$}

Now we proceed to the generalization of the ideas mentioned above
by looking at a more general ansatz allowing string dynamics on
three angles in $S^5$ part of the geometry. In this section we
will closely follow the notation of \cite{AFZ} which we find more
suitable for our ansatz than the method based on the Complex
sine-Gordon model used for the analysis of the two spin case in
\cite{Dorey2}. The background of interest is \footnote{Different
multispin string configurations on $\axs$ and their connections
with known integrable models were studied in \cite{arts}. In
\cite{ryang} some revant multi spin rotating string configurations
were analyzed.}:

\eq{ ds^2=-dt^2+d\theta^2+\cos^2\th
d\phi_3^2+\sin^2\th\left(d\psi^2+\cos^2\psi d\phi_1^2+\sin^2\psi
d\phi_2^2\right) \label{2.18}}

Now we will define a new coordinate $z=\sin\theta$, fix
$\psi=\pi/4$ and work in conformal gauge $t=\tau$. The Polyakov
action takes the form:

\eq{\begin{array}{l} S_P=-\ds\frac{\sqrt{\l}}{4\pi}\int d\s d\tau
[-\dot{t}^2+\ds\frac{1}{1-z^2}((\p_{\s}z)^2-(\p_{\tau}z)^2)+(1-z^2)((\p_{\s}\phi_3)^2-(\p_{\tau}\phi_3)^2)\\\\
\ds\frac{z^2}{2}\left((\p_{\s}\phi_1)^2-(\p_{\tau}\phi_1)^2+(\p_{\s}\phi_2)^2-(\p_{\tau}\phi_2)^2\right)]
\end{array}\label{2.19}}

We should impose also the Virasoro constraints:

\eq{\begin{array}{l}
\ds\frac{\dot{z}z'}{1-z^2}+(1-z^2)\dot{\phi_3}\phi_3'+z^2(\dot{\phi_1}\phi_1'+\dot{\phi_2}\phi_2')=0\\\\
\ds\frac{\dot{z}^2+z'^2}{1-z^2}+(1-z^2)(\dot{\phi_3}^2+\phi_3'^2)+z^2(\dot{\phi_1}^2+\phi_1'^2+\dot{\phi_1}^2+\phi_2'^2)=1
\end{array}\label{2.20}}

It can be shown that the following anzatz is consistent with the
equations of motion and the above constraints:

\eq{ z=z(\s-v\tau) \qquad \phi=\tau+g(\s-v\tau) \qquad
\phi_1=\o_1\tau-v\o_1\s \qquad \phi_2=\o_2\tau-v\o_2\s
\label{2.21}}

Since we have fixed $\psi=\pi/4$ the equation of motion for $\psi$
imposes the constraint $\o_1=\o_2=\o_{12}$ \footnote{We thank the
authors of \cite{kruruts} for pointing out an error in the first
version of this paper.}. Plugging this into the Virasoro
constraints leads to (we use $y=\s-v\tau$):

\eq{\begin{array}{l} \p_yg=\ds\frac{vz^2}{(1-v^2)(1-z^2)}\\\\
(z')^2=\ds\frac{z^2}{(1-v^2)^2}[(1-v^2)(1-\o_{12}^2(1-v^2))-(1-\o_{12}^2(1-v^2)^2)z^2]
\end{array}\label{2.22}}

For future convenience we define:

\eq{
\xi=\ds\sqrt{\ds\frac{(1-v^2)(1-\o_{12}^2(1-v^2))}{1-\o_{12}^2(1-v^2)^2}}
\qquad \g=\ds\sqrt{\frac{1-\o_{12}^2(1-v^2)}{1-v^2}} \label{2.23}}

We can integrate the equation for $z(y)$ and find:

\eq{ z=\ds\frac{\xi}{\cosh(\g y)}\label{2.24}}

Now we present the expressions for the quantities of interest:

\eq{\begin{array}{l} p=\ds\int_{-r}^{r}d\s\phi_3' \qquad
E=\ds\frac{\sqrt{\l}}{2\pi}\ds\int_{-r}^{r}d\s \qquad
J_3=\ds\frac{\sqrt{\l}}{2\pi}\ds\int_{-r}^{r}d\s(1-z^2)\dot{\phi_3}\\\\
J_1=\ds\frac{\sqrt{\l}}{2\pi}\ds\int_{-r}^{r}d\s
\ds\frac{z^2}{2}\dot{\phi_1} \qquad
J_1\ds\frac{\sqrt{\l}}{2\pi}\ds\int_{-r}^{r}d\s \ds\frac{z^2}{2}
\dot{\phi_2}
\end{array}\label{2.25}}

Using the solution we have for $z(y)$ we can solve this integrals
and find (we take $r\rightarrow \infty$):

\eq{\begin{array}{l} E-J_3=\ds\frac{\l\xi^2}{\pi\g(1-v^2)} \qquad
\xi=\sin\left(\frac{p}{2}\right)\\\\
J_1=\ds\frac{\l\xi^2}{2\pi\g}\o_{12} \qquad
J_2=\ds\frac{\l\xi^2}{2\pi\g}\o_{12}
\end{array}
\label{2.26}}

Now using the relation:

\eq{\o_{12}^2+\ds\frac{\g^2}{\xi^2}=\ds\frac{1}{(1-v^2)^2}
\label{2.27}}

We find the dispersion relation for the two spin giant magnon:

\eq{
E-J_3=\ds\sqrt{(J_1+J_2)^2+\ds\frac{\l}{\pi^2}\sin^2\left(\ds\frac{p}{2}\right)}
\label{2.28}}

This is a dispersion relation which reproduces the one for giant
magnons with one and two spins found in \cite{HM} and
\cite{Dorey1}. We see that the dispersion relation for our simple
two spin giant magnon anzatz effectively reduces to the 2 spin
dispersion relation (\ref{2.22}) if we identify $J_3\rightarrow
J_1, \,\, J_1+J_2\rightarrow J_2$. From spin chain side one can
think of this result as it was discussed in \cite{Dorey1} for two
spin bound states (see discussion after eq. (16)). Therefore the
result can be interpreted as superposition of two wavepackets -
one in $\vf_1$ another in $\vf_2$ direction. Thus we see that our
solution is a particular case of the more general 3 spin giant
magnon found in \cite{kruruts}. The solution we found somehow
looks similar to the situation in the magnon excitations in the
deformed $\axs$ considered in \cite{khoze} - to obtain the single
spin magnon dispersion relation one should involve dynamics on
another angle from five sphere, while in the limit $\b\rightarrow
0$ it still reproduces single magnon case.


\sect{$\b$-deformed $\axs$}

Here we review the $\b$-deformed $\axs$ background found by Lunin
and Maldacena \cite{LM}. This background is conjectured to be dual
to marginal deformations of $\N=4$ SYM. We note that this
background can be obtained from pure $\axs$ by a series of TsT
transformations as described in \cite{Frolov}. Here we will
restrict ourselves to the case of real deformation parameter. The
resulting supergravity background for string theory dual of real
$\b$-deformations of $\N=4$ SYM is:

\eq{
ds^2=R^2\left(ds^2_{AdS_5}+\ds\sum_{i=1}^{3}(d\m_i^2+G\m_i^2d\phi_i^2)+\t{\g}^2G\m_1^2\m_2^2\m_3^2(\ds\sum_{i=1}^3d\phi_i^2)\right)
\label{3.1}}

This background includes also a dilaton field as well as RR and
NS-NS form fields. The relevant form for our classical string
analysis will be the antisymmetric B-field:

\eq{
B=R^2\t{\g}G\left(\m_1^2\m_2^2d\phi_1d\phi_2+\m_2^2\m_3^2d\phi_2d\phi_3+\m_1^2\m_3^2d\phi_1d\phi_3\right)
\label{3.2}}

In the above formulas we have defined

\eq{\begin{array}{l} \t{\g}=R^2\g \qquad\qquad R^2=\sqrt{4\pi
g_sN}=\sqrt{\l}
\\\\
G=\ds\frac{1}{1+\t{\g}^2(\m_1^2\m_2^2+\m_2^2\m_3^2+\m_1^2\m_3^2)}\\\\
\m_1=\sin\theta\cos\a \qquad \m_2=\cos\theta \qquad
\m_3=\sin\theta\sin\a
\end{array}\label{3.3}}

Where ($\theta$,$\a$,$\phi_1$,$\phi_2$,$\phi_3$) are the usual
$S^5$ variables and the deformation parameter is $\b=\g+i\s_d$ is
complex in general, but in our analysis we will consider $\s_d=0$


\subsection{Two spin giant magnons in $\b$-deformed $\axs$}

We will follow \cite{khoze} and consider semiclassical strings on
the $R\times S^3$ part of the deformed Lunin-Maldacena background.
The metric and the antisymmetric NS-NS field are:

\eq{ ds^2=-dt^2+d\theta^2+G\sin^2\theta d\phi_1^2+G\cos^2\theta
d\phi_2^2\label{3.4}}

\eq{ B=\t{\g} G \sin^2\theta\cos^2\theta\label{3.5}}

where

\eq{G=\ds\frac{1}{1+\t{\g}^2\sin^2\theta\cos^2\theta}\label{3.6}}

The Polyakov action takes the form:

\eq{\begin{array}{l} S_P=\ds\frac{\sqrt{\l}}{4\pi}\int d\t d\s
[-(\p_{\tau}t)^2-(\p_{\tau}\theta)^2+(\p_{\s}\theta)^2+G\sin^2\theta((\p_{\s}\phi_1)^2-(\p_{\tau}\phi_1)^2)+\\\\
G\cos^2\theta((\p_{\s}\phi_2)^2-(\p_{\tau}\phi_2)^2)-2\t{\g}
G\sin^2\theta\cos^2\theta(\p_{\tau}\phi_2\p_{\s}\phi_1-\p_{\tau}\phi_1\p_{\s}\phi_2)]
\end{array}\label{3.7}}

We define the variable $y=cx-dt$ and require $c^2-d^2=1$ such that
we always have $v=d/c\leq 1$. Now we will investigate the
following rotating string anzatz (which is consistent with the
equation of motion and the two Virasoro constraints).

\eq{\theta=\theta(y) \qquad \phi_1=t+g_1(y) \qquad \phi_2=\n
t+g_2(y)\label{3.8}}

This anzatz was studied in \cite{AFZ} in the pure $\axs$
background and describes a giant magnon with two spins, one of
which is going to infinity and the other is finite. By using the
equations of motion derived from the Polyakov action and the two
Virasoto constraints we find:

\eq{\begin{array}{l}
\p_yg_1=\ds\frac{d}{G\sin^2\theta}-d-c\n\t{\g}\cos^2\theta\\\\
\p_yg_2=-\n d-c\t{\g}\sin^2\theta\\\\
\p_y\theta=\cos\theta\ds\sqrt{1-(\n c+\t{\g}
d)^2-d^2\frac{\cos^2\theta}{\sin^2\theta}}
\end{array}\label{3.9}}

We can integrate the last of these equations and find:

\eq{ \cos\theta=\ds\frac{\cos\theta_0}{\cosh(\sqrt{1-(\n c+\t{\g}
d)^2}y)} \label{3.10}}

where we have defined

\eq{\cos\theta_0=\ds\sqrt{\frac{1-(\n c+\t{\g} d)^2}{c^2-(\n
c+\t{\g} d)^2}}\label{3.11}}

Following \cite{HM} and \cite{khoze} we will identify

\eq{\cos\theta_0=\sin(\frac{p}{2}-\pi\b)\label{3.12}}

where $p$ is the momentum of the magnon and $\b$ is the
deformation parameter (in our case $\b$ is real and
$\t{\g}=\sqrt{\l}\b$). Now it is easy to find the conserved
charges in our problem ($J_1$ and $J_2$ correspond to $\phi_1$ and
$\phi_2$ respectively):

\eq{\begin{array}{l}
E=\ds\frac{\sqrt{\l}}{\pi}\ds\int_{-\infty}^{\infty}d\s\\\\
J_1=\ds\frac{\sqrt{\l}}{\pi}\ds\int_{-\infty}^{\infty}d\s
G\sin^2\theta\left((1-d\p_yg_1)-\t{\g}\cos^2\theta\p_yg_2\right)\\\\
J_2=\ds\frac{\sqrt{\l}}{\pi}\ds\int_{-\infty}^{\infty}d\s
G\cos^2\theta\left((\n-d\p_yg_2)+\t{\g}\sin^2\theta\p_yg_1\right)
\end{array}\label{3.13}}

Now using our solution for $\theta(y)$ we find:

\eq{E-J_1=\ds\frac{\sqrt{\l}c}{\pi}\ds\frac{\cos^2\theta_0}{\sqrt{1-(\n
c+\t{\g} d)^2}} \label{3.14}}

and

\eq{ J_2=\ds\frac{\n c+\t{\g} d}{c}(E-J_1) \label{3.15}}

From here we find

\eq{(E-J_1)^2=J_2^2+\ds\frac{\l}{\pi^2}\cos^2\theta_0
\label{3.16}}

So finally we find the two spin magnon dispersion relation:

\eq{
E-J_1=\sqrt{J_2^2+\ds\frac{\l}{\pi^2}\sin^2(\frac{p}{2}-\pi\b)}
\label{3.17}}

We note that when the deformation parameter $\b$ vanishes we
reproduce the dispersion relation (\ref{2.17}) for the two spin
magnon in pure $\axs$ found. The above result also agrees with the
conclusions of \cite{Beisert:2005if} that in the deformed theory
we should modify the asymptotic Bethe ansatz equations by the
transformation $p \rightarrow p-2\pi\b$. As expected from the
Hofman-Maldacena interpretation \footnote{We thank S. Pal for some
comments on this point} the dispersion relation is periodic in $p$
with period $2\pi$ and in $\b$ with period $1$ .


\sect{Conclusions and Outlook}

In this short paper we investigated the multispin string solutions
corresponding to giant magnon excitations. First we give a simple
generalization of the two spin string solutions corresponding to
magnon excitations. Our solution with two spins in the spherical
part is the same as the giant magnons with two spins but in our
case string dynamics on another isometry of the sphere is alowed.
Although the solution is similar, it extends the slass of string
solutions corresponding to two spin magnon excitations. As a
second example we considered the giant magnon solution with two
spins in the $\b$-deformed $\axs$ background. Our solution looks
identical to the one found in \cite{khoze}, but in our case the
solution is "essentially" with two spins in the following sense.
While in \cite{khoze} the solution reproduces the one spin
solution in the limit $\gamma\rightarrow 0$, in our case this
limit reproduces the two spin $\axs$ solution obtained in
\cite{AFZ},\cite{MTT}.

In conclusion, the Hofman-Maldacena solutions certainly deserves
more detailed investigations. One direction for further study is
to include more spins from the AdS part of the background, by
considering classical strings moving on $AdS_3\times S^3$ and
$AdS_3\times S^5$ and thus generalizing the solution found in
\cite{MTT} for strings moving on $AdS_3\times S^1$. It would be
interesting also to find the corresponding spin chain dispersion
relations. Another possibly interesting problem would be to see
how the giant magnon solutions look like in the non-supersymmetric
multiparameter $\b_i$-deformed background found by Frolov
\cite{Frolov} and further studied in
\cite{tseytfrroi},\cite{Prinsloo:2005dq},\cite{Alday:2005ww}. Some
of these issues are currently under investigation.

\vspace{1.cm}

\newpage
{\large \textit{Note}:}

After the completion of this paper two interesting papers
\cite{volovich},\cite{kruruts} discussing multispin giant magnons
appeared. The authors of \cite{volovich} analyze different
scattering and bound states of one and two spin giant magnons
using a technique based on the dressing method for the $SO(N)$
vector model, in the last section of this paper they present a
particular three spin solution. It is interesting that the dressing method allows more general time dependence beyond particular ansatz\footnote{We thank M. Spradlin for pointing this to us.}. In \cite{kruruts} a 
three spin giant magnon solution was found by reductions to Neumann-Rosochatius integrable system. Along with other giant
magnon solutions and a thorough analysis on the spin chain side is given.

\vspace{0.5cm} {\large\textbf{Acknowledgements}}

We thank M. Kruczenski, J. Russo and A.A. Tseytlin for pointing
out a mistake in the first version of our paper. We thank A.A.
Tseytlin for usefull comments.  The work of N.P.B. is supported by
a University of Southern California Graduate Research Fellowship.
The work of R.R is partially supported by a Bulgarian NSF
CNVUY-10/66 grant.

\end{document}